\documentstyle[preprint,aps]{revtex}
\begin{document}
\title{
Implications of Recent Data \\
on Non-mesonic Decay of Light $\Lambda$-hypernuclei}

\author{Namyoung Lee$^a$, Hong Jung$^b$, and Dongwoo Cha$^a$}

\address{
$^a$
Department of Physics, Inha University, Inchon 402-751, KOREA}

\address{
$^b$
Department of Physics, Sookmyung Women's University, Seoul 140-742, 
KOREA}

\maketitle

\begin{abstract}
We analyze the recent data on the non-mesonic decays of light 
$\Lambda$-hypernuclei up to ${}^{12}_{\Lambda}{\rm C}$
using the phenomenological model of Block and Dalitz.
Fitting the spin-isospin dependent $\Lambda N\to NN$ reaction rates
to six data points, we predict the remaining data in reasonable 
consistency.
We find that despite the short-range nature of the $\Lambda N\to NN$
interaction, the non-mesonic decay of $p$-shell hypernuclei seems to 
be strongly induced by the $p$-shell neutrons. 
Also, the recent data indicate that 
the $\Delta I= \frac{1}{2}$ rule,
well-proved at the hadronic level, may not be sacred in the nuclear
medium and the $\Delta I = \frac{3}{2}$ interactions seem to be 
needed to describe the non-mesonic decays of $\Lambda$-hypernuclei. 
\end{abstract}


\newpage

The structure of hypernuclei has been understood somewhat 
successfully in terms of empirical hyperon-nucleon potentials and 
mean field theories as in the case of ordinary nuclei\cite{Bando}. 
However, the decay rates of hypernuclei have revealed intriguing 
features\cite{Bando,BD,Cohen,Schumacher}.
The $\Lambda$-hypernucleus, whose ground state is stable with respect 
to strong interactions, decays only via weak interactions and so far 
$\Lambda\to N\pi$ (so called pi-mesonic decay) and $\Lambda N\to
NN$ (two-body non-mesonic decay) have been considered 
as its main decay mechanisms although there have been some reports on 
the importance of the three-body non-mesonic process 
$\Lambda NN\to NNN$ \cite{LambdaNN}.

The pi-mesonic decays ($\Lambda\to p\pi^-$ and $\Lambda\to n\pi^0$) 
account for most of the decays of free $\Lambda$-hyperons whose decay 
width is 
\begin{equation}
\Gamma_{\Lambda}\cong 2.50\mu {\rm eV}. 
\label{free-L}
\end{equation}
In the rest frame of the $\Lambda$,
the energy release is $m_{\Lambda}-m_N-m_{\pi}\cong 35$ MeV and
the momentum of the emitted nucleon is about $100$ MeV.
Also, the ratio of the decay widths for the two channels is
$\Gamma(\Lambda\to p\pi^-)/ \Gamma(\Lambda\to n\pi^0)\approx 1.8$,
which means that $\Delta I= \frac{1}{2}$ amplitude dominates over the 
$\Delta I= \frac{3}{2}$ amplitude as is in the free decays of 
$K$-mesons. This is the well-known $\Delta I= \frac{1}{2}$ rule, 
although it is yet to be understood why it should hold true at the 
hadronic level.

When the $\Lambda$-hyperon is bound in the nuclear medium, its 
pi-mesonic decay mode becomes  suppressed because the allowed phase 
space is smaller due to the small energy release from the bound 
$\Lambda$-hyperon and the emitted nucleons are Pauli-blocked. 
Therefore, the non-mesonic decay mode whose energy release is about 
$m_{\Lambda}-m_N\cong 175$ MeV, much larger than in the pi-mesonic 
case, becomes more important as the hypernucleus becomes heavier. 
However, contrary to the pi-mesonic decay for which we have a 
well-known effective Hamiltonian satisfying the 
$\Delta I= \frac{1}{2}$ rule, 
we do not have such a reliable effective Hamiltonian for the 
non-mesonic decay of $\Lambda$-hypernuclei yet.  The main reason is
that there are no experimental data on the free 
$\Lambda N\to \Lambda N$ reaction. So far most of the theoretical 
studies have been based on meson-exchange models with the 
$\Delta I= \frac{1}{2}$ rule assumed. However, their predictions for
the partial decay widths, $\Gamma_n$ for $\Lambda n\to nn$ and
$\Gamma_p$  for $\Lambda p\to np$, are in general not quite 
consistent with the  experimental data, although they are strongly 
model dependent \cite{Bando,Oset}.

Decades ago, Block and Dalitz\cite{BD} analyzed the measurements of 
the non-mesonic decays of ${}^4_{\Lambda}{\rm H}$ and ${}^4_{\Lambda}
{\rm He}$ in terms of the spin and isospin dependence of the 
$\Lambda N\to NN$ interaction without referring to any specific 
effective Hamiltonian.
Assuming that the interaction is local and the $\Lambda$ decay 
induced by different nucleons is incoherent, they expressed the
non-mesonic decay rates of hypernuclei in terms of the elementary 
reaction rates for $\Lambda N\to NN$ with definite angular momentum 
and the mean nucleon density at the $\Lambda$ position.
By fitting the elementary reaction rates to the available empirical 
data of hypernuclear decays, one could do meaningful predictions on 
other non-mesonic decay rates of hypernuclei.

Based on the phenomenological approach of Block and Dalitz, 
Cohen\cite{Cohen} and Schumacher\cite{Schumacher} tried to determine 
the elementary decay rates using the data on ${}^4_{\Lambda}{\rm H}$, 
${}^4_{\Lambda}{\rm He}$, and ${}^5_{\Lambda}{\rm He}$. 
However, Cohen used the same old data on ${}^4_{\Lambda}{\rm H}$ and 
${}^4_{\Lambda}{\rm He}$ as in Ref.\cite{BD}, and Schumacher used the 
controversial preliminary data on ${}^4_{\Lambda}{\rm He}$ from the 
Brookhaven E-788 experiment. 
Now that we have more reliable experimental data on 
${}^4_{\Lambda}{\rm H}$\cite{Outa1}, 
${}^4_{\Lambda}{\rm He}$\cite{Outa2}, 
${}^5_{\Lambda}{\rm He}$\cite{BNL1}, 
${}^{11}_{\Lambda}{\rm B}$ and ${}^{12}_{\Lambda}{\rm C}$
\cite{BNL1,KEK}, we are in a better position to do theoretical
investigations of non-mesonic decays of light $\Lambda$-hypernuclei.
In this work, we reanalyze the $s$-shell hypernuclei 
(${}^4_{\Lambda}{\rm H}$, ${}^4_{\Lambda}{\rm He}$,
${}^5_{\Lambda}{\rm He}$) using the recent data and also extend 
the approach of Block and Dalitz to $p$-shell hypernuclei 
(${}^{11}_{\Lambda}{\rm B}$, ${}^{12}_{\Lambda}{\rm C}$).

According to Block and Dalitz\cite{BD}, the non-mesonic decay width 
of hypernucleus ${}^A_{\Lambda}{\rm Z}$ may be written as
\begin{equation}
  \Gamma_{nm}({}^A_{\Lambda} {\rm Z}) = \rho ({}^A_{\Lambda} 
  {\rm Z}){\bar R},
\end{equation}
where $\bar R$ is the spin and isospin average of the elementary 
reaction rates $R_{NJ}$'s for $\Lambda N\to NN$ ($N=p$, $n$) with the 
total angular momentum of the $\Lambda N$ pair being $J$, and 
$\rho({}^A_{\Lambda}{\rm Z})$ is the mean nucleon density at the 
$\Lambda$ position.
$\rho({}^A_{\Lambda}{\rm Z})$ is given by
\begin{equation}
  \rho({}^A_{\Lambda}{\rm Z})= 
  (A-1)\int d^3{\vec r}\rho_N({\vec r})|\psi_{\Lambda}({\vec r})|^2,
\label{rho}
\end{equation}
with $\rho_N$ and $\psi_{\Lambda}$ being the normalized nucleon 
density  and the $\Lambda$ wavefunction in the rest frame of the 
nuclear core, respectively. 
In this model, it is straightforward to write down the non-mesonic 
decay widths of light $\Lambda$-hypernuclei in terms of the four 
$R_{NJ}$'s \cite{BD,Cohen}:
\begin{eqnarray}
  \Gamma_{nm}({}^4_{\Lambda}{\rm H})  &=&
   {1\over 6}\rho({}^4_{\Lambda}{\rm H})(2R_{p0}+3R_{n1}+R_{n0}) 
            \label{4H} \\
  \Gamma_{nm}({}^4_{\Lambda}{\rm He}) &=&
   {1\over 6}\rho({}^4_{\Lambda}{\rm He})(3R_{p1}+R_{p0}+2R_{n0}) \\
  \Gamma_{nm}({}^5_{\Lambda}{\rm He}) &=&
   {1\over 8}\rho({}^5_{\Lambda}{\rm He})(3R_{p1}+R_{p0}+3R_{n1}+
    R_{n0}) 
\end{eqnarray}
where all the nucleons are in the $s$-shell and the total angular 
momenta of hypernuclei are 0, 0, and $\frac{1}{2}$ for 
${}^4_{\Lambda}{\rm H}$, ${}^4_{\Lambda}{\rm He}$, and 
${}^5_{\Lambda}{\rm He}$, respectively. Here, the $\Lambda$-hyperon 
is treated as being in the  $s$-state because the decay of
hypernuclei is expected to occur in  their ground states.

In order to apply this model to the $p$-shell hypernuclei, we 
need to take into account the possibility of $\Lambda N\to NN$ 
reaction with the initial $\Lambda N$ pair with relative orbital 
angular momentum of one, that is, the non-mesonic decay induced by 
the $p$-shell nucleons. 
Thus, in general,  we need to introduce
six more elementary rates, $R_{NJ}'$ for $J=0$, $1$, and $2$. 
For ${}^{11}_{\Lambda}{\rm B}$ and ${}^{12}_{\Lambda}{\rm C}$, 
however, the total angular momenta are $\frac{5}{2}$ and $1$, 
respectively, 
so that the $\Lambda N$ pair can have total angular momentum of one 
or two and there appear only two spin averaged elementary rates 
$R_N'\equiv (3R_{N1}'+5R_{N2}')/8$ with $N=p, n$.
To be specific, the non-mesonic decay widths of 
${}^{11}_{\Lambda}{\rm B}$ and ${}^{12}_{\Lambda}{\rm C}$
are given by
\begin{eqnarray}
  \Gamma_{nm}({}^{11}_{\Lambda}{\rm B}) =
    {1\over 8}\rho_s({}^{11}_{\Lambda}{\rm B})(3R_{p1} &+&R_{p0}+
       3R_{n1}+R_{n0})+ {1\over 2}\rho_p({}^{11}_{\Lambda}{\rm B})
           (R_{p}'+R_{n}') \\
  \Gamma_{nm}({}^{12}_{\Lambda}{\rm C}) =
      {1\over 8}\rho_s({}^{12}_{\Lambda}{\rm C})
        (3R_{p1} &+& R_{p0}+3R_{n1}+R_{n0}) 
      + {1\over 7}\rho_p({}^{12}_{\Lambda}{\rm C})
          (4R_{p}'+3R_{n}') 
\label{12C}
\end{eqnarray}
where $\rho_s$ ($\rho_p$) represents the mean $s$ state ($p$ state) 
nucleon density at the $\Lambda$ position.

In the mean nucleon densities $\rho({}^A_{\Lambda}{\rm Z})$ of  
Eq.(\ref{rho}), the nucleon wavefunction is obtained using a 
Woods-Saxon potential with its parameters adjusted to give the 
correct r.m.s. radius of the nucleus ${}^{A-1}Z$ and the $\Lambda$ 
wavefunction is similarly obtained by fitting the $\Lambda$ binding 
energy of the hypernucleus ${}^A_{\Lambda}Z$. For the $p$-shell 
hypernuclei, we take the spin-orbit interaction into account as in 
Ref.\cite{Bando}.
The resulting mean nucleon densities at the $\Lambda$ position are 
given in the second row of Table \ref{tab1}.

The six elementary decay rates
$R_{p0}$, $R_{p1}$, $R_{n0}$, $R_{n1}$, $R_p'$ and $R_n'$ 
can be determined uniquely from the relations similar to 
Eqs. (\ref{4H}) - (\ref{12C}) provided that six reliable 
data points are available. As for such data points, we take the 
three total non-mesonic decay widths of ${}^4_{\Lambda}{\rm H}$, 
${}^5_{\Lambda}{\rm He}$, and ${}^{11}_{\Lambda}{\rm B}$, and 
the two proton-induced partial decay widths of 
${}^5_{\Lambda}{\rm He}$ and ${}^{11}_{\Lambda}{\rm B}$, and
finally the ratio $\Gamma_n / \Gamma_p$ of 
${}^4_{\Lambda}{\rm He}$. 
In solving the simultaneous equations, we
took the mean values of the data points, the bold-faced letters in 
the last three rows of Table \ref{tab1}. 

Our estimated values for the elementary decay rates $R_{NJ}$'s 
are summarized in Table \ref{tab2}. Since we have determined all six 
of the necessary elementary decay rates of the model, we are now in 
a position to predict other decay widths. Unfortunately, however, 
we are left with only three experimental data to compare with 
up to now, namely $\Gamma_{nm}({}^{12}_{\Lambda}{\rm C})$, 
$\Gamma_p({}^{12}_{\Lambda}{\rm C})$ and 
$\Gamma_{nm}({}^4_{\Lambda}{\rm He})$. Our predictions for
these decay widths are given under the heading of `Our model' 
together with the measured values in Table \ref{tab1}. It can be
seen from the table  that our predicted widths fall nicely not only
within error bounds of the data but also within a few percents of 
the mean values except for the ${}^4_{\Lambda}{\rm He}$ case where 
there is still some  controversy on the measured decay width. 
Therefore, we conclude  with reasonable confidence that the 
phenomenological model used  in our analysis seems to work fairly 
well for light $\Lambda$-hypernuclei.

Once we admit that the model establishes the elementary decay rates
$R_{NJ}$'s reliably from the adopted data, we can find two
interesting implications from our numerical results for the six 
$R_{NJ}$'s shown in  Table \ref{tab2}. Firstly, $R_n'$, the 
spin-averaged elementary decay rate induced by the $p$-shell 
neutrons, turns out to be relatively quite large. This indicates 
that the non-mesonic $\Lambda$-hyperon decay induced by  the 
$p$-shell nucleons (particularly by the neutrons) is not less 
probable than the decay induced by the $s$-shell nucleons. 
Certainly, this result is in contradiction
to a naive expectation from the short-range nature of the 
$\Lambda N \to NN$ interaction that the main contribution to the
decay rate should come from the $\Lambda N$ pair with zero orbital
angular momentum\cite{Fetisov}.  In the present situation without a
working microscopic mechanism for  the non-mesonic $\Lambda$-hyperon 
decay, it remains to be  understood how the $p$-shell nucleons
contribute to the decay of $\Lambda$-hypernuclei. 
In particular, the three-body non-mesonic process
$\Lambda nn\to nnn$\cite{LambdaNN} may compete with the two body 
process $\Lambda n\to nn$ induced by the $p$-shell neutrons. 
Secondly, if the
$\Delta I= \frac{1}{2}$ rule should hold true in nuclear medium, we 
would have\cite{BD}
\begin{equation}
   {R_{n0} \over R_{p0}} =2, \hspace{0.25in}
   {R_{n1} \over R_{p1}} \le 2,  \hspace{0.25in}
          {R_n'\over R_p'}\le 2, \hspace{0.25in} 
\label{Delta-oh}
\end{equation}
while if $\Delta I= \frac{3}{2}$ should hold true,  we would
have \cite{Schumacher} 
\begin{equation}
  { R_{n0} \over R_{p0}} = \frac{1}{2}. \hspace{0.5in}
\label{Delta-th}
\end{equation}
Now, according to our results given in Table \ref{tab2}, we have
\begin{equation}
  {R_{n0} \over R_{p0} } \cong 0.4, \hspace{0.25in}
  {R_{n1} \over R_{p1} } \cong 1.5, \hspace{0.25in}
  {R_n' \over R_p'} \cong 11.5, \hspace{0.25in}
  {\rm (our \;\; model \;\; prediction)}.
\label{Delta-our}
\end{equation}
Comparing the predicted elementary reaction rates of ours in 
Eq.(\ref{Delta-our}) with those of Eq. (\ref{Delta-oh}) 
and Eq. (\ref{Delta-th})
it is evident that the adopted 
recent data favor including $\Delta I = \frac{3}{2}$ interactions 
in addition to $\Delta I = \frac{1}{2}$ ones in the effective 
Hamiltonian for the non-mesonic decay of $\Lambda$-hypernuclei. 
However, with the experimental error bars taken into 
account, one may argue that the 
$\Delta I= \frac{1}{2}$ rule is not  
excluded completely. Nevertheless, it seems quite plausible
that $\Delta I= \frac{3}{2}$ amplitude can be as important as the
$\Delta I= \frac{1}{2}$ amplitude in the non-mesonic decays of
$\Lambda$-hypernuclei. Based on the conclusions of the present work,
a microscopic approach using an effective Hamiltonian with 
four-fermion interactions of $\Delta I= \frac{1}{2}$
and $\Delta I=\frac{3}{2}$ in addition to the 
$\Delta I= \frac{1}{2}$ pion-exchange interactions is in progress. 
 
In summary, we have analyzed the recent data on the non-mesonic 
decays of the $s$-shell hypernuclei (${}^4_{\Lambda}{\rm H}$, 
${}^4_{\Lambda}{\rm He}$, ${}^5_{\Lambda}{\rm He}$) using the 
phenomenological model of Block and Dalitz and also extended our 
analysis to $p$-shell hypernuclei (${}^{11}_{\Lambda}{\rm B}$, 
${}^{12}_{\Lambda}{\rm C}$). According to the model, the 
non-mesonic decay widths of all the $s$-shell as well as the 
$p$-shell $\Lambda$-hypernuclei can be expressed in terms of only 
six of the elementary reaction rates $R_{NJ}$'s. Making use of the 
six data points from recently measured decay widths, we can 
determine the six elementary reaction rates uniquely. They are used  
in turn to predict the decay widths of the remaining data 
to check our model calculations. The results are certainly more than 
satisfactory even though there exist only three available decay 
widths to compare with which have been measured up to now.
{}From solely the six elementary reaction rates determined by the 
recent data, we can deduce two interesting implications. One is that 
despite the short-range nature of the $\Lambda N \to NN$ 
interaction, the non-mesonic decay of $p$-shell hypernuclei seems 
to be strongly induced by the $p$-shell neutrons. The other is that 
the recent data favor including $\Delta I= \frac{3}{2}$ interactions 
in addition to $\Delta I = \frac{1}{2}$ ones in the effective 
Hamiltonian for the non-mesonic decay of $\Lambda$-hypernuclei. 
Therefore, we suggest that the $\Delta I= \frac{1}{2}$ rule,  
well-proved at the hadronic level, may not be sacred in the nuclear 
medium. 

We are grateful to Hyoung Chan Bhang for the information regarding 
the present status of the data on ${}^4_{\Lambda}{\rm He}$ through 
his private communication with H. Outa and for valuable discussions. 
We also thank Jung-Hwan Jun for useful comments. The work of H.J. 
was supported in part by the Korea Science and Engineering 
Foundation through the Center for Theoretical Physics of Seoul 
National University and in part by the 1996 research grant of 
Sookmyung Women's University.
The work of D.C. was supported in part by a research grant from 
Inha University in 1994, and by the Ministry of Education, Korea 
(BSRI-95-2424).

\begin{table}
\caption
{Input data for our model calculation and results from our model
calculations. The second row is the mean nucleon density at the
$\Lambda$-hyperon position calculated with the usual Woods-Saxon
potential. From the third to the fifth rows, measured non-mesonic
decay widths of light $\Lambda$-hypernuclei are given. Data points
used in determining the elementary reaction rates are written in bold
face numbers. `Our Model' represents decay widths predicted by
our calculations.} 
\begin{tabular}{cccccc}
  $\Lambda$-hypernucleus    & ${}^4_\Lambda {\rm H}$
& ${}^4_\Lambda {\rm He}$   & ${}^5_\Lambda {\rm He}$ 
& ${}^{11}_\Lambda {\rm B}$   & ${}^{12}_\Lambda {\rm C}$ \\
\tableline
  Mean nucleon density & $\rho =$0.017    & $\rho =$0.019
& $\rho=$0.038         & $\rho_s =$0.048   & $\rho_s =$0.047 \\
  ( ${\rm fm}^{-3}$ )   &  &  &  &  
  $\rho_p =$ 0.038     & $\rho_p =$ 0.044 \\
\tableline
  $\Gamma_{nm} / \Gamma_\Lambda $   
& {\bf 0.24} $\pm$ 0.13 \tablenotemark[1]
& 0.17 $\pm$ 0.05 \tablenotemark[2]
& {\bf 0.41} $\pm$ 0.14 \tablenotemark[3] 
& {\bf 0.95} $\pm$ 0.17 \tablenotemark[4]
& 0.89 $\pm$ 0.18 \tablenotemark[4]  \\
  Our Model   &  &   0.20  &  &  &   0.95   \\
\tableline
  $\Gamma_n / \Gamma_p$  &  
& {\bf 0.40} $\pm$ 0.15 \tablenotemark[5]
&  &  &  \\
\tableline
  $\Gamma_p / \Gamma_\Lambda$  &  &  
& {\bf 0.21} $\pm$ 0.07 \tablenotemark[3]
& {\bf 0.30} ${}_{-0.11}^{+0.15}$ \tablenotemark[4]
& 0.31 ${}_{-0.11}^{+0.18}$ \tablenotemark[4] \\
  Our Model  &  &  &  &  &   0.31 \\
\end{tabular}
\tablenotetext[1]{Taken from Ref. \cite{Outa1}.}
\tablenotetext[2]{Taken from Ref. \cite{Outa2}.}
\tablenotetext[3]{Taken from Ref. \cite{BNL1}.}
\tablenotetext[4]{Taken from Ref. \cite{KEK}.}
\tablenotetext[5]{Taken from Ref. \cite{Block}.}
\label{tab1}
\end{table}
\begin{table}
\caption{The elementary reaction rates in the unit of 
$\Gamma_\Lambda {\rm fm}^3 $ determined from six 
data points written by bold face letters in Table \ref{tab1}.}
\begin{tabular}{cccc}
\multicolumn{2}{c}{$s$-shell}   &  \multicolumn{2}{c}{$p$-shell} \\
\cline{1-2} 
\cline{3-4}
proton  &  neutron  &  proton  &  neutron  \\
\tableline
$R_{p0}=$21.3 & $R_{n0}=$8.8  & $R_p '=$1.8 & $R_n '=$20.9 \\
$R_{p1}=$7.6  & $R_{n1}=$11.1 &  & \\
\end{tabular}
\label{tab2}
\end{table}

\end{document}